\documentclass[usenatbib,usegraphicx]{mn2e}
\usepackage{amsfonts}
\usepackage{amsmath}
\usepackage{amssymb}
\usepackage{comment}
\usepackage{color}

\title[Minor-merger-driven star formation]
    {The significant contribution of minor mergers to the cosmic star formation budget \vspace{-0.2in}}

\author[Sugata Kaviraj]
{S. Kaviraj\\
Centre for Astrophysics Research, University of Hertfordshire,
College Lane, Hatfield, Herts, AL10 9AB, UK\\\vspace{-0.3in}}

\begin{document}

\maketitle

\def \aj {AJ}
\def \mnras {MNRAS}
\def \pasp {PASP}
\def \apj {ApJ}
\def \apjs {ApJS}
\def \apjl {ApJL}
\def \aap {A\&A}
\def \nat {Nature}
\def \araa {ARAA}
\def \iaucirc {IAUC}
\def \aaps {A\&A Suppl.}
\def \qjras {QJRAS}
\def \na {New Astronomy}
\def \aapr {A\&ARv}
\def\lesssim{\mathrel{\hbox{\rlap{\hbox{\lower4pt\hbox{$\sim$}}}\hbox{$<$}}}}
\def\gtrsim{\mathrel{\hbox{\rlap{\hbox{\lower4pt\hbox{$\sim$}}}\hbox{$>$}}}}


\begin{abstract}
We estimate an empirical lower limit for the fraction of cosmic
star formation that is triggered by minor mergers in the local
Universe. Splitting the star-formation budget by galaxy
morphology, we find that early-type galaxies (ETGs) host
$\sim$14\% of the budget, while Sb/Sc galaxies host the bulk
($\sim$53\%) of the local star formation activity. Recent work
indicates that star formation in nearby ETGs is driven by minor
mergers, implying that at least $\sim$14\% of local star formation
is triggered by this process. A more accurate estimate can be
derived by noting that an infalling satellite likely induces a
larger starburst in a galaxy of `later' morphological type, both
due to higher availability of gas in the accreting galaxy and
because a bigger bulge better stabilizes the disk against star
formation. This enables us to use the star formation in ETGs to
estimate a lower limit for the fraction of star formation in
late-type galaxies (LTGs) that is minor-merger-driven. Using a
subsample of ETGs that is mass and environment-matched to the LTGs
(implying a similar infalling satellite population), we estimate
this limit to be $\sim$24\%. Thus, a \emph{lower limit} for the
fraction of \emph{cosmic} star formation that is induced by minor
mergers is $\sim$35\% (14\% [ETGs] + 0.24$\times$86\% [LTGs]).
{\color{black}The observed positive correlation} between black
hole and galaxy mass further implies that a similar fraction of
black hole accretion may also be triggered by minor mergers.
Detailed studies of minor-merger remnants are therefore essential,
to quantify the role of this important process in driving stellar
mass and black hole growth in the local Universe.
\end{abstract}




\section{Introduction}
Understanding the processes that drive stellar mass growth is a
central topic in observational cosmology. While the galaxy
evolution literature has traditionally relied on relatively small
samples of galaxies \citep[e.g.][]{deZeeuw2002}, the advent of
large spectro-photometric surveys, like the Sloan Digital Sky
Survey (SDSS), is revolutionizing our understanding of the
statistical properties of galaxy populations, by delivering
samples of objects numbering in the hundreds of thousands.

A topic that is rapidly gaining prominence is the significant role
of minor mergers (mass ratios $\lesssim$ 1:4) in influencing the
evolution of massive galaxies
{\color{black}\citep[e.g.][]{Weinmann2009,Nierenberg2011,Wang2012}}.
Recent work on nearby early-type (elliptical and lenticular)
galaxies that reside in {\color{black}low-density environments
(groups and the field)} indicates that the star formation in these
systems is driven by minor mergers. {\color{black}In the
\emph{HST-COSMOS} field -- which, by virtue of being an `empty'
field, is dominated by low-density environments -- a strong
correlation is observed between morphological disturbances in ETGs
and blue UV colours (due to young stars). This indicates that the
star formation is driven by mergers and not by processes like
stellar mass loss or accretion that would leave the stellar
morphology unperturbed \citep{Kaviraj2011}}. However, the
major-merger rate at late epochs ($z<1$) is far too low to satisfy
the number of disturbed ETGs, indicating that \emph{minor} mergers
drive the star formation in these objects \citep{Kaviraj2011}.

Theoretical work supports this observational picture. At late
epochs the gas supply from internal stellar mass loss is an order
of magnitude too low to produce the levels of star formation
implied by the blue UV colours of star-forming ETGs
\citep[][]{Kaviraj2007c}. Furthermore, \citet{Kaviraj2009} have
combined realistic numerical simulations of minor mergers with the
expected frequency of these events in the $\Lambda$CDM paradigm,
to show that the predicted UV-optical colour distributions are
remarkably consistent with observations (see their Fig 3). The
relative lack of native sources of gas makes ETGs excellent
tracers of minor-merger-driven star formation. The star formation
is observed against a `blank canvas' of old stars, enabling us to
quantify the impact of this process with good precision.

While our focus in this study is on star formation, it is worth
noting that the work on minor-merger-driven stellar mass growth is
mirrored by studies of the size evolution of massive ETGs. Both
observational \citep[e.g.][]{Trujillo2006,Huertas-Company2012} and
theoretical
{\color{black}\citep[e.g.][]{Bournaud2007,Naab2009,Oser2012}} work
suggests that minor mergers are likely responsible for the bulk of
the factor 3-5 evolution observed in ETG sizes over cosmic time
\citep[e.g.][]{Daddi2005,Cimatti2012}.

While it is increasingly clear that minor mergers may strongly
influence both the star-formation and structural properties of
massive galaxies, their total contribution to the cosmic star
formation budget remains poorly understood. Quantifying this
contribution holds the key to understanding the \emph{overall}
significance of this process in influencing galaxy evolution, and
is the principal aim of this paper.

This Letter is organized as follows. In Section 2, we describe the
galaxy sample -- drawn from the SDSS Stripe 82 -- that underpins
this study, the classification of galaxy morphologies via visual
inspection and the measurement of galaxy properties such as
stellar masses and star formation rates (SFRs). In Section 3 we
explore the cosmic stellar mass and star formation budgets in the
local Universe and in Section 4 we quantify the minor-merger
contribution to local star formation. We summarize our findings in
Section 5. {\color{black}Throughout, we use the \emph{WMAP3}
cosmological parameters \citep{Komatsu2011}: $\Omega_m = 0.241$,
$\Omega_{\Lambda} = 0.759$, $h=0.732$, $\sigma_8 = 0.761$.}

\begin{figure}
\begin{center}
$\begin{array}{c}
\includegraphics[width=2.35in]{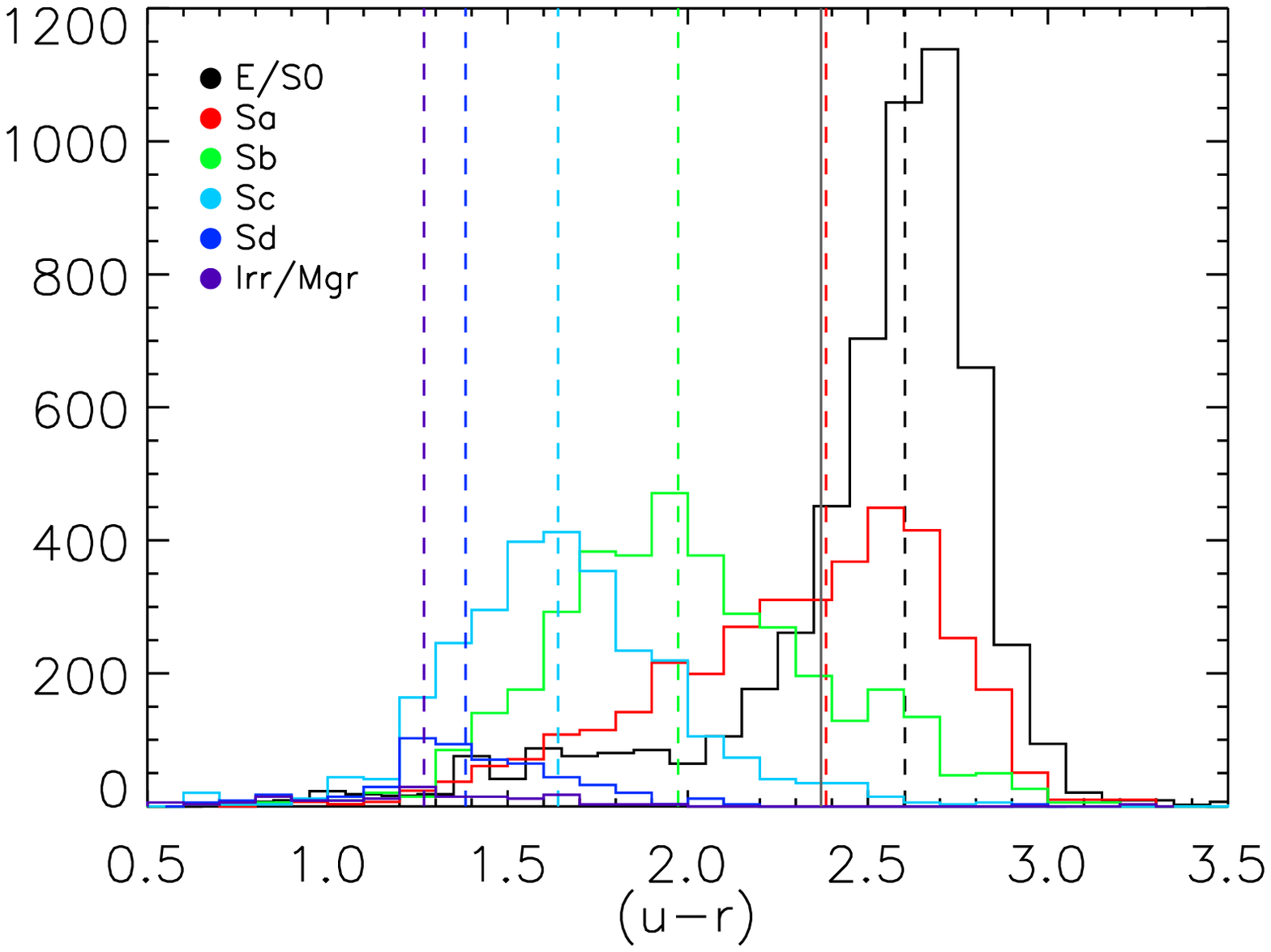}\\
\includegraphics[width=2.35in]{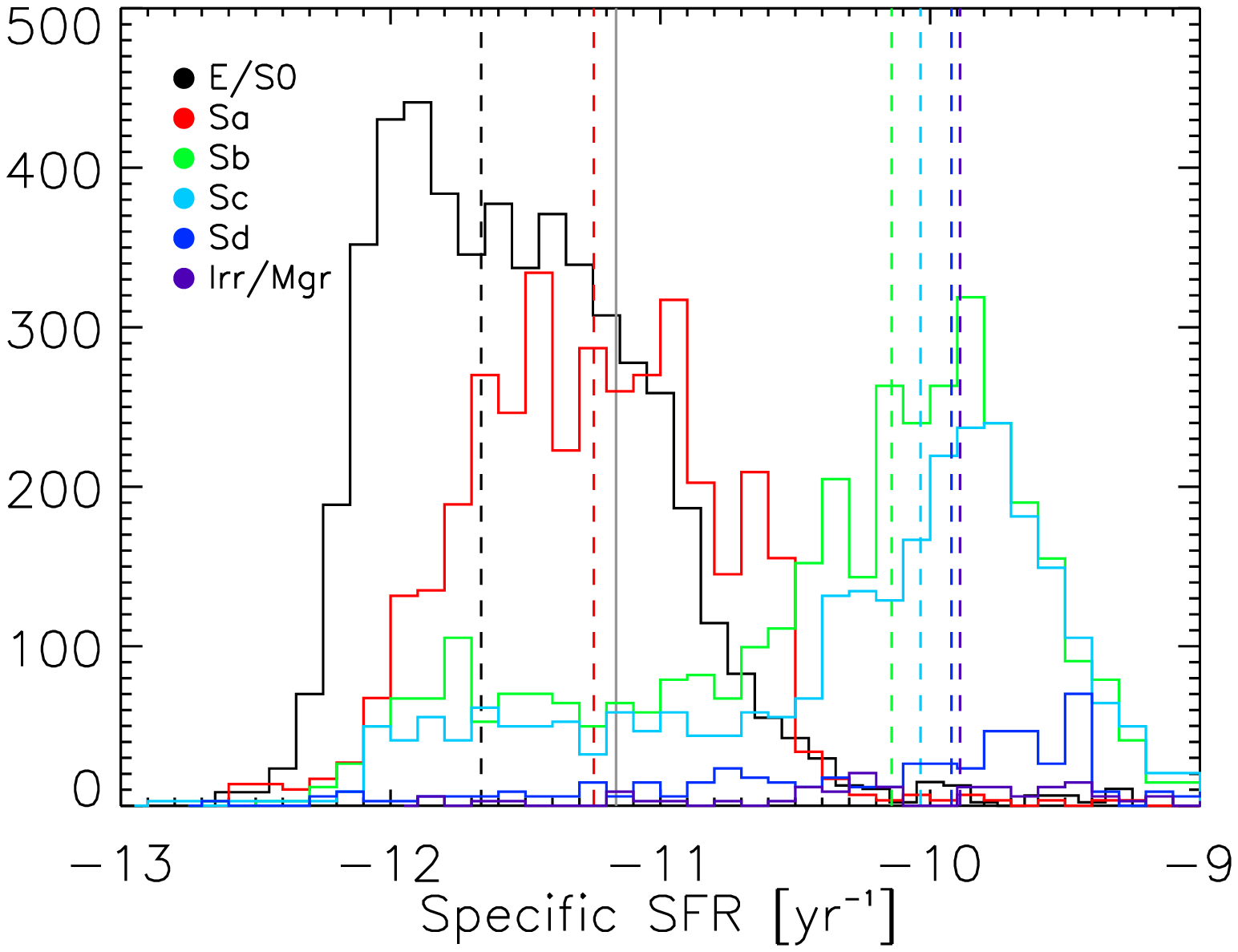}
\end{array}$
\caption{Rest-frame $(u-r)$ colours (top) and specific SFRs
(bottom) of galaxies of various morphological types (see legends).
Dashed vertical lines indicate median values. {\color{black}The
solid grey line indicates a sample of ETGs that is matched in
stellar mass and environment to the LTGs (see Section 4)}. The
magnitudes are K-corrected using the publicly available
\texttt{KCORRECT} code of \citet{Blanton2007}.} \label{fig:u_r}
\end{center}
\end{figure}


\vspace{-0.15in}
\section{Data}
\subsection{Galaxy sample and morphological classification}
The galaxies in this study are drawn from the SDSS Stripe 82, a
region along the celestial equator in the Southern Galactic Cap
$(-50^{\circ} < \alpha < 59^{\circ}, -1.25^{\circ} < \delta <
1.25^{\circ})$ that has been multiply imaged as part of the SDSS
Supernova Survey \citep{Frieman2008}. The $\sim300$ deg$^2$ area
of the Stripe 82 offers a co-addition of 122 imaging runs
\citep{Abazajian2009}, yielding images that are $\sim$2 mags
deeper than the standard-depth, 54 second SDSS scans (which have
magnitude limits of $22.2$, $22.2$ and $21.3$ mags in the $g$, $r$
and $i$-bands respectively).

We classify galaxy morphologies via visual inspection of both
standard-depth, multi-colour images from the SDSS DR7 and their
deeper $r$-band Stripe 82 counterparts. Recent work has widely
used visual inspection for morphological classification of bright
($r<16.8$), nearby ($z<0.1$) SDSS galaxies \citep[see
e.g.][]{Kaviraj2007c,Lintott2008,Nair2010}. Here we restrict our
study to $r<16.8$ and $z<0.07$, which yields a sample $\sim$6,500
galaxies. {\color{black}Note that all galaxies in this sample have
stellar mass, SFR and environment measurements, on which our
subsequent analysis is based.}

While past studies have typically not exploited the deep Stripe 82
imaging for morphological classification, these images enhance our
ability to identify faint disks and tidal features that are
difficult to detect in the standard-depth SDSS exposures,
maximising the accuracy of the morphological classification
\citep{Kaviraj2010}. We classify galaxies into standard
morphological types \citep{Hubble1926,dev1959}: E, S0, Sa, Sb, Sc,
Sd, Irregular/Mergers. The percentage of galaxies in our
morphological classes are E : S0 : Sa : Sb : Sc : Sd : Irr/Mgr =
19.5$^{\pm 0.84}$ : 21.5$^{\pm 0.88}$ : 18.1$^{\pm 0.81}$ :
21.4$^{\pm 0.88}$ : 15.1$^{\pm 0.74}$ : 3.39$^{\pm 0.35}$ :
1.01$^{\pm 0.22}$. This is similar to equivalent values from past
studies, such as Fukugita et al. (2007, see their Fig 13): 20.6 :
24.5 : 16.6 : 19.8 : 15.0 : 2.10 : 1:30. The slightly higher
fraction of late-type systems in our study, compared to previous
work based purely on standard-depth images, is plausibly driven by
better identification of faint disks in the deeper Stripe 82
scans.

\begin{figure}
  \begin{minipage}{0.5\textwidth}
    \begin{center}
        \includegraphics[width=2.8in]{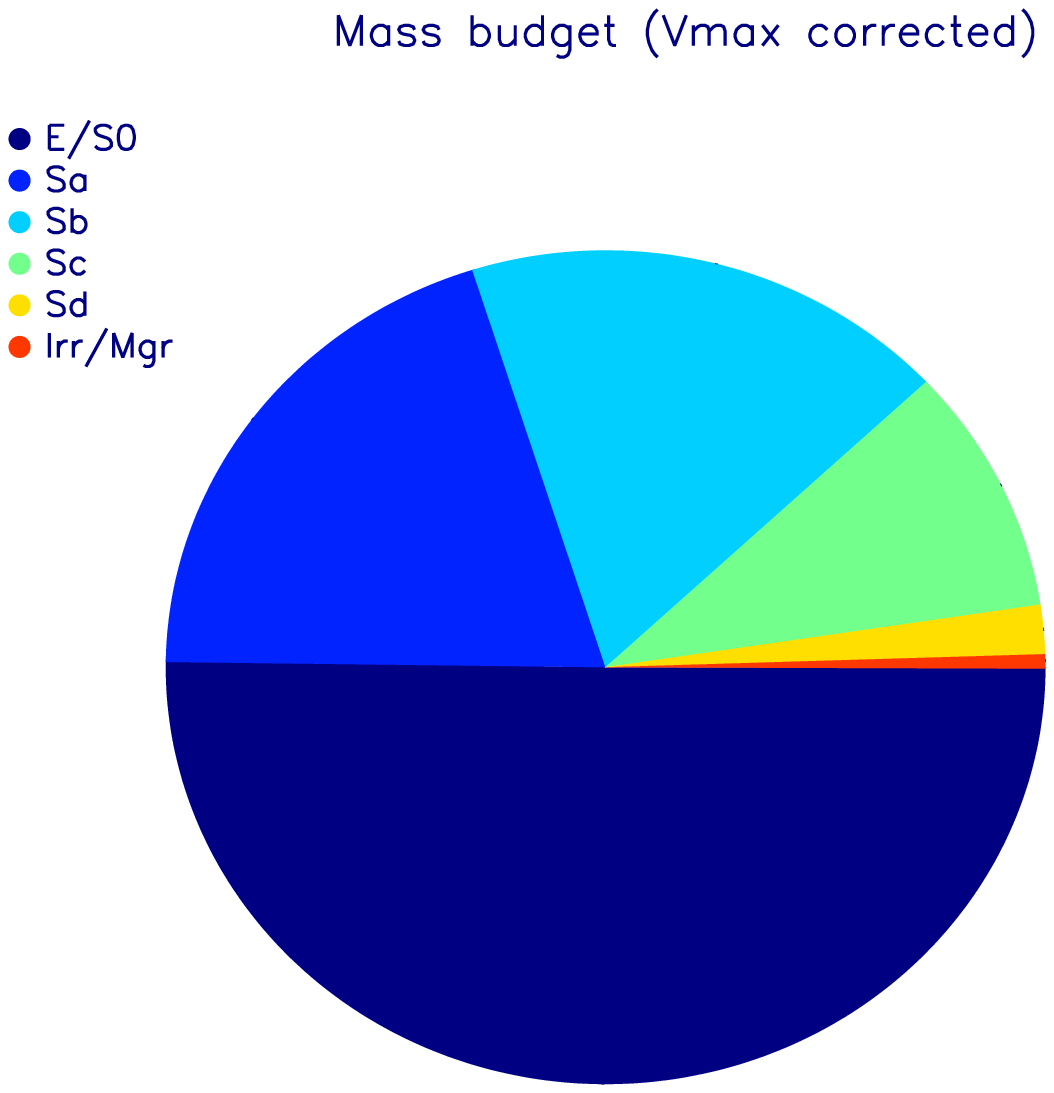}
    \end{center}
  \end{minipage}\vspace{-0.2in}
  \begin{minipage}{0.5\textwidth}
     \begin{center}
        \begin{tabular}{l|c}

            \textbf{Morphology} & \textbf{Fraction of stellar mass budget}\\
            E/S0    & 0.502$^{\pm 0.068}$\\
            Sa      & 0.199$^{\pm 0.043}$\\
            Sb      & 0.179$^{\pm 0.039}$\\
            Sc      & 0.096$^{\pm 0.020}$\\
            Sd      & 0.019$^{\pm 0.003}$\\
            Irr/Mgr & 0.005$^{\pm 0.001}$

        \end{tabular}
    \end{center}
  \caption{The cosmic stellar mass budget at $z<0.07$, split by galaxy morphology and corrected for Malmquist
  bias. {\color{black}The errors are calculated by combining the uncertainties on individual galaxies using standard error propagation formulas.}}
\end{minipage}
\label{fig:mass_budget}
\end{figure}


\begin{figure}
  \begin{minipage}{0.5\textwidth}
    \begin{center}
        \includegraphics[width=2.8in]{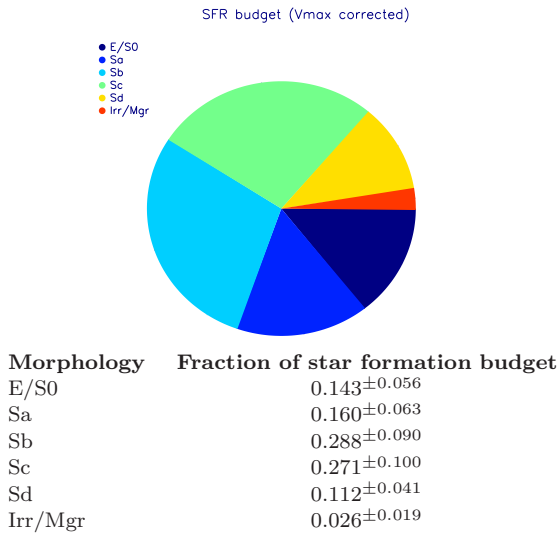}
    \end{center}
  \end{minipage}\vspace{-0.2in}
  \begin{minipage}{0.5\textwidth}
     \begin{center}
        \begin{tabular}{l|c}

            \textbf{Morphology} & \textbf{Fraction of star formation budget}\\
            E/S0    & 0.143$^{\pm 0.056}$\\
            Sa      & 0.160$^{\pm 0.063}$\\
            Sb      & 0.288$^{\pm 0.090}$\\
            Sc      & 0.271$^{\pm 0.100}$\\
            Sd      & 0.112$^{\pm 0.041}$\\
            Irr/Mgr & 0.026$^{\pm 0.019}$

        \end{tabular}
    \end{center}
  \caption{The cosmic star formation budget at $z<0.07$, split by galaxy morphology and corrected for Malmquist bias. {\color{black}The errors are calculated by combining the uncertainties on individual galaxies using standard error propagation formulas.}}
\end{minipage}
\label{fig:sfr_budget}
\end{figure}


\subsection{Stellar masses, star formation rates and local environments}
{\color{black}We employ published stellar masses
\citep{Kauffmann2003} and star formation rates \citep[SFRs;][B04
hereafter]{Brinchmann2004} from the latest (DR7) version of the
publicly-available MPA-JHU SDSS catalogue
\footnote{http://www.mpa-garching.mpg.de/SDSS/DR7/}, which is the
largest, most homogeneous value-added SDSS catalogue to date.}
{\color{black}Briefly, stellar masses are calculated by comparing
galaxy $ugriz$ photometry to a large grid of synthetic star
formation histories, based on the \cite{BC2003} stellar models and
assuming a \citet{Kroupa2001} initial mass function.} Model
likelihoods are calculated from the values of $\chi^2$, and 1D
probability distributions for free parameters like stellar mass
are constructed via marginalisation. The median of the 1D
distribution provides a best estimate for the parameter in
question, with the 16th and 84th percentile values (which enclose
68\% of the total probability) yielding an associated `1$\sigma$'
uncertainty. {\color{black}The median stellar mass of our galaxies
is $\sim$10$^{10.3}$ M$_{\odot}$ and the 5th and 95th percentile
values of the mass distribution are $\sim$10$^{8}$ M$_{\odot}$ and
$\sim$10$^{11.2}$ M$_{\odot}$ respectively (i.e. 90\% of the
galaxies lie between these two values).}

SFRs are estimated via two different methods
\citep[see][]{Brinchmann2004}, depending on the classification of
the galaxy in a standard optical emission-line ratio analysis
\citep[][see also Baldwin et al. 1981; Kewley et al.
2006]{Kauffmann2003}. For galaxies classified as `star-forming'
(i.e. where the nuclear emission is dominated by star formation),
SFRs are estimated by comparing galaxy spectra to a library of
models based on \citet{Charlot2001}, with a dust treatment that
follows the empirical model of \citet{Charlot2000}. For galaxies
that are unclassified (due to their emission lines being too weak)
or that are classified as `AGN/Composite' (in which a significant
fraction of the nuclear emission is from a central AGN), SFRs are
calculated using the D4000 break. {\color{black}The D4000 feature
is produced via blanket absorption of short-wavelength photons
from metals in stellar atmospheres, and is stronger in galaxies
that are deficient in hot, blue stars
\citep[e.g.][]{Poggianti1997}. This makes it a useful estimator of
SFR in systems with weak emission lines, or those in which the
diagnostic lines are contaminated by non-thermal sources (e.g.
AGN)}. The relation between specific SFR and D4000 for the
`star-forming' subsample is used to estimate the specific SFR of
the galaxy in question, which is then converted to a total SFR via
the measured galaxy stellar mass. Note that the SFRs used here are
both extinction and aperture-corrected. {\color{black}55\% of our
galaxies are classified as `star-forming', with 20\% unclassified
and 25\% classified as AGN/Composite. The median SFR of our
galaxies is $\sim$1 M$_{\odot}$ yr$^{-1}$ and the 5th and 95th
percentile values of the SFR distribution are $\sim$0.4
M$_{\odot}$ yr$^{-1}$ and $\sim$6.8 M$_{\odot}$ yr$^{-1}$
respectively (i.e. 90\% of the galaxies lie between these two
values). It is worth noting that the emission-line results in B04
are consistent with the recent literature
\citep[e.g.][]{Tresse1998,Perez2008} and their D4000-measured SFRs
for ETGs (which have a median value of $\sim$0.08 M$_{\odot}$
yr$^{-1}$) are in agreement with those calculated via
UV/optical/IR photometry \citep[e.g.][]{Kaviraj2013b}.}

When considering the cosmic star formation and stellar mass
budgets below, we must correct for the fact that brighter galaxies
in our magnitude-limited sample can be observed to larger
distances (Malmquist bias). We perform this correction via the
standard method of weighting galaxy properties by the maximum
volume over which they can be detected (V$_{\textnormal{MAX}}$; Mo
et al. (2010)), given the magnitude limit of the SDSS
spectroscopic sample ($r=17.77$ mag).

Finally, to estimate galaxy environment we employ the environment
catalogue of \citet{Yang2007}, who use an iterative halo-based
group finder to separate the SDSS into structures across a broad
dynamical range, spanning clusters to isolated galaxies. The
catalogue provides estimates of the mass of the dark-matter (DM)
halo that hosts each SDSS galaxy, which can be used as a proxy for
environment \citep[e.g.][]{vdb2002}. For example `cluster-sized'
halos are typically more massive than $10^{14}$M$_{\odot}$,
`group-sized' halos have masses between $10^{13}$M$_{\odot}$ and
$10^{14}$M$_{\odot}$, while smaller halos constitute what is
commonly termed the `field', e.g. \citet{Kaviraj2009c}).


\vspace{-0.15in}
\section{The cosmic star formation and stellar mass budgets}
Figure \ref{fig:u_r} presents rest-frame $(u-r)$ colour
distributions (top) and specific SFRs (bottom) for different
morphological classes \citep[see also][]{Strateva2001}. The dashed
vertical lines indicate median values. Not unexpectedly, the
colours become bluer as galaxies move towards `later'
morphological types. {\color{black}In Figures 2 and 3} we present
the (V$_{\textnormal{MAX}}$ corrected) stellar mass and star
formation rate budgets. In agreement with previous work
\citep[e.g.][]{Bernardi2009}, ETGs account for almost half the
stellar mass budget in the local Universe. Indeed, systems with
prominent bulges (ETGs + Sa systems) account for $\sim$70\% of the
stellar mass in today's Universe, with disk-dominated galaxies (Sb
and later morphological types) hosting the remaining $\sim$30\%.
In contrast, ETGs contribute only $\sim$14\% of the cosmic star
formation budget, which is dominated by late-type systems, with Sb
and Sc galaxies hosting the bulk ($\sim$53\%) of the star
formation activity in the local Universe.


\section{The minor-merger contribution to the star formation
budget} {\color{black}We note first that, by virtue of being drawn
from a blind, wide area survey, our galaxies predominantly inhabit
low-density environments (groups and the field, e.g. Kaviraj
2010). Moreover, since environmental processes such as
ram-pressure stripping and harassment make clusters hostile to
star formation \citep[e.g.][]{Dressler1984,Moore1999,Kimm2011},
cluster galaxies are not expected to contribute significantly to
the cosmic star formation budget, leaving our overall results
unaffected.}

If star formation in ETGs is driven by minor mergers, then the ETG
portion of the star formation budget ($\sim$14\%) places an
absolute lower limit on the minor-merger contribution to local
star formation. However, since all systems, regardless of
morphology, experience minor mergers, calculating the overall
contribution of this process also requires an estimate of the
minor-merger-driven fraction of star formation in late-type
galaxies (LTGs).

It is reasonable to expect that, all other things being equal
(galaxy mass, environment etc.), the same satellite falling into a
system with a `later' morphological type will trigger a larger
starburst. This is both because of the availability of more gas,
since later-type systems have higher native gas fractions
\citep[e.g.][]{Kannappan2004}, but also because a larger bulge
better stabilizes the gas disk against star formation
\citep[e.g.][]{Hernquist1995,Martig2009}. Simulations of minor
mergers \citep[e.g.][]{Hernquist1989,Mihos1994,Hernquist1995}
indicate that this is indeed the case. Minor mergers induce radial
inflows of native disk gas (as well as gas brought in by the
satellite itself), triggering starbursts in the central regions of
the primary galaxy. However, in a simulated minor merger with a
1:10 mass ratio, the star formation produced by an identical
satellite is almost an order of magnitude lower if the system
hosts a modest bulge that has 1/3 of the mass of the disk
\citep[e.g.][]{Mihos1994}. Not unexpectedly, disks with larger gas
fractions and less prominent bulges experience larger gas inflows
and central star formation, given the same initial merger
configuration \citep[][]{Hernquist1995}. Since ETGs, by
definition, have the most prominent bulges and negligible native
gas reservoirs compared to late-type galaxies, we can use the star
formation activity in ETGs to calculate a lower limit for the
minor-merger-driven activity in their late-type counterparts.

However, we note first that ETGs and LTGs typically have different
distributions in stellar mass and local environment
\citep[e.g.][]{Kauffmann2003}, both of which will affect the
characteristics of the infalling satellite population. For
example, more massive galaxies will attract more satellites, while
infalling satellites are expected to be more gas-poor in denser
environments, due to processes like ram-pressure stripping
\citep[e.g.][]{Kimm2011}. {\color{black}Therefore, we first draw
an ETG subsample that has the same stellar mass and environment
distribution as the LTGs (indicated by the solid grey lines in
Figure \ref{fig:u_r}). This ETG subset contains $\sim$1300
galaxies, with a median stellar mass, SFR and $(u-r)$ colour of
10$^{10.3}$ M$_{\odot}$, $\sim$0.2 M$_{\odot}$yr$^{-1}$ and
$\sim$2.4 mag. Compared to the full ETG sample, this subset is a
factor of $\sim$2 less massive, with a median SFR that is a factor
of $\sim$2 higher. Finally, while this ETG subset is matched in
mass and environment to the LTGs, its median SFR is around a
factor of $\sim$3 lower than that of the LTGs.}

From a statistical point of view, this ETG subsample will
experience a similar infalling satellite population to that
experienced by the LTGs. We divide the total star formation rate
in this ETG subsample ($\psi^{tot}_{ETG'}$) by its total stellar
mass ($M^{tot}_{ETG'}$), which yields an estimate for the
(globally-averaged) star formation activity per unit mass in these
ETGs. We then multiply this value by the total stellar mass in the
LTGs ($M^{tot}_{LTG}$), to produce an estimate for the star
formation that is attributable to the minor-merger process
($\psi^{MM}_{LTG}$) in the late-type galaxy population. In other
words,

\begin{equation}
\psi^{MM}_{LTG} = \frac{\psi^{tot}_{ETG'}}{M^{tot}_{ETG'}} \times
M^{tot}_{LTG}.
\end{equation}

If similar satellites typically produce more young stars in
galaxies of later morphological type, then
$\psi^{\textnormal{LTG}}_{\textnormal{MM}}$ becomes a reasonably
robust lower limit to the star formation in the LTG population
that is plausibly triggered by minor mergers. Our calculated value
for $\psi^{\textnormal{MM}}_{\textnormal{LTG}}$ is $\sim24$\% of
the total star formation budget hosted by the LTGs. In other words
\emph{at least} a quarter of the star formation activity in
today's LTG population is likely triggered by minor mergers. Since
late-type systems host $\sim$86\% of the overall star formation
budget, this then implies that a \emph{lower limit} for the
fraction of \emph{cosmic} star formation that is induced by minor
mergers locally is then $\sim$35\% (14\% [ETGs] + 0.24$\times$86\%
[LTGs]).

{\color{black}It is worth noting that the observed correlation
between stellar and black-hole mass in galaxies of all
morphological types \citep[see e.g.][]{Gultekin2009,McConnell2011}
indicates that, on average, stellar mass buildup and accretion on
to the central black hole occur in lockstep. This, in turn,
implies that a similar fraction of black hole accretion in today's
Universe is also likely to be induced by minor mergers. Our
results therefore indicate that \emph{minor mergers drive at least
a significant minority ($\sim$35\%) of local star formation and
black hole accretion} that should not be ignored.}

Persistent star formation in ETGs at $z<1$ \citep{Kaviraj2008},
coupled with the paucity of major mergers at late epochs
\citep[e.g.][]{Stewart2008,Lopez2009,Conselice2009}, indicates
that minor mergers remain influential in driving star formation
over at least the latter half of cosmic time \citep{Kaviraj2011}.
Indeed, minor-merger-driven star formation is likely to become
more intense at progressively higher redshift
\citep[e.g.][]{Kaviraj2008}, due to the higher gas fractions in
both the infalling satellites and the accreting galaxy. While the
$z>1$ Universe remains comparatively unexplored, both
observational \citep[e.g.][]{ForsterSchreiber2011} and theoretical
\citep[e.g.][]{Keres2009,Dekel2009a} work indicates that major
mergers are an insignificant driver of cosmic star formation at
these epochs, hosting only $\sim$15\% of the cosmic star formation
budget at $z \sim2$ \citep{Kaviraj2013a}. This implies a
significant role for other processes such as minor mergers and
direct accretion (e.g. via cold streams) in building the old stars
that dominate today's Universe
\citep[e.g.][]{Dekel2009b,Kaviraj2013a}. Indeed, the triggering of
starbursts by accreted satellites is likely to have been more
efficient in the early Universe if disks formed before bulges.
{\color{black}Systematic empirical studies of minor-merger-driven
stellar-mass and black-hole growth are therefore essential, both
in the nearby Universe and at high redshift.}


\section{Summary}
We have estimated an empirical \emph{lower limit} to the fraction
of cosmic star formation that is likely to be triggered by minor
mergers in the local Universe, using a sample of $\sim$6,500
bright ($r<16.8$), nearby ($z<0.07$) galaxies in the SDSS Stripe
82. We have split our galaxy sample into standard morphological
classes (E, S0, Sa, Sb, Sc, Sd, Irr/Merger), via visual inspection
of both their standard-depth multi-colour images and their deeper
$r$-band Stripe 82 counterparts. The availability of deep imaging
enables us to better identify faint disks and tidal features,
maximizing the accuracy of the visual inspection. The percentages
of galaxies in our morphological classes are E : S0 : Sa : Sb : Sc
: Sd : Irr/Mgr = 20.6$^{\pm 2.30}$ : 21.7$^{\pm 2.30}$ :
18.8$^{\pm 2.30}$ : 20.8$^{\pm 2.30}$ : 14.4$^{\pm 2.30}$ :
2.89$^{\pm 2.30}$ : 0.62$^{\pm 2.30}$.

{\color{black}In agreement with previous work
\citep{Bernardi2009}}, we have found that ETGs account for around
half the stellar mass budget in the local Universe, with
disk-dominated galaxies (Sb and later morphological types)
accounting for $\sim$30\%. In contrast, ETGs contribute only
$\sim$14\% of the cosmic star formation budget, while Sb and Sc
galaxies host the bulk ($\sim$53\%) of the local star formation
activity.

{\color{black} Previous work has demonstrated that star formation
in nearby ETGs that inhabit low-density environments is driven by
minor mergers \citep{Kaviraj2011}}. The ETG portion of the star
formation budget ($\sim$14\%) is therefore an absolute lower limit
to the minor-merger contribution to local star formation. However,
since all galaxies, regardless of morphology, experience minor
mergers, a more accurate calculation requires an estimate of the
fraction of star formation in LTGs that is also attributable to
this process.

It is likely that, all other things being equal (galaxy mass,
environment, etc.), the same satellite will trigger a larger
starburst in a system of `later' morphological type, both due to
greater availability of native gas and because a larger bulge
better stabilizes the disk against star formation (a picture
supported by numerical simulations of minor mergers). This allows
us to use the star formation in ETGs to estimate a lower limit for
the fraction of star formation in LTGs that is likely to be
induced by minor mergers.

Using a subsample of ETGs that is mass and environment-matched to
the LTGs (implying similar infalling satellite populations) we
have estimated that this lower limit to the minor-merger-driven
fraction of star formation in local LTGs is $\sim$24\%. Our
results then imply that a lower limit to the fraction of
\emph{cosmic} star formation that is induced by minor mergers is
$\sim$35\% (14\% [ETGs] + 0.24$\times$86\% [LTGs]). The observed
correlation between black hole and galaxy mass further suggests
that a similar fraction of the black hole accretion in the local
Universe may also be triggered by minor mergers. Thus minor
mergers are likely to drive a significant minority of star
formation and black hole accretion in the local Universe.

Recent work indicates that minor mergers remain influential in
fuelling star formation in ETGs at least until $z\sim1$. Together
with an emerging literature which suggests that the major-merger
channel triggers a relatively insignificant fraction of cosmic
star formation at $z>1$, this suggests that minor mergers may have
had an important role in stimulating the growth of a large
fraction of the stellar mass in today's Universe. Given our poor
current understanding of the minor-merger process, detailed
observational and theoretical studies of minor-merger remnants are
essential, in order to fully quantify the significant role of this
process in galaxy evolution.


\vspace{-0.2in}
\section*{Acknowledgements}
I thank the referee for many useful comments which have improved
the paper. Martin Hardcastle, Timothy Davis and Martin Bureau are
thanked for interesting discussions.


\vspace{-0.2in}
\nocite{Baldwin1981,Kewley2006,Mo2010}

\small
\bibliographystyle{mn2e}
\bibliography{references}


\end{document}